# AN INTEGER GARCH MODEL FOR A POISSON PROCESS WITH TIME VARYING ZERO-INFLATION

Isuru Ratnayake[1*] and V.A. Samaranayake[2]

1. Department of Biostatistics and Data Science, Kansas University Medical Center, Kansas City, KS.
2. Department of Mathematics and Statistics, Missouri University of Science and Technology, Rolla, MO.

## ABSTRACT

A time-varying zero-inflated serially dependent Poisson process is proposed. The model assumes that the intensity of the Poisson Process evolves according to a generalized autoregressive conditional heteroscedastic (GARCH) formulation. The proposed model is a generalization of the zero-inflated Poisson Integer GARCH model proposed by Fukang Zhu in 2012, which in return is a generalization of the Integer GARCH (INGARCH) model introduced by Ferland, Latour, and Oraichi in 2006. The proposed model builds on previous work by allowing the zero-inflation parameter to vary over time, governed by a deterministic function or by an exogenous variable. Both the Expectation Maximization (EM) and the Maximum Likelihood Estimation (MLE) approaches are presented as possible estimation methods. A simulation study shows that both parameter estimation methods provide good estimates. Applications to two real-life data sets show that the proposed INGARCH model provides a better fit than the traditional zero-inflated INGARCH model in the cases considered.

*Corresponding author E-mail address: rratnayake@ku.edu

# 1. INTRODUCTION

The standard Poisson point process, which assumes statistical independence between observations, is not suitable for modeling time series of counts that display serial dependence. One way to address this deficiency is to define a time series where the count at a given time is generated by a Poisson distribution whose intensity parameter is dependent on past counts and past intensities. Observe that for a discrete time process defined in such a manner with a linear dependent structure and observed at equally spaced time points, the intensity parameter at a given time is the mean count at that time conditioned on the past information. Rydberg and Shephard (2000) proposed such a model, where the current conditional mean is a linear function of both the observed count and the conditional mean at the pervious time point. When encountering such models, where the condition mean of the Poisson distribution is the intensity parameter, we will refer to it as the conditional mean rather than intensity unless the use of the later term is warranted for clarity. Similar models have been proposed by several other authors and these are discussed in Chapter 4 of the book by Kedem and Fokianos (2002). Andreas Heinen (2003) generalized the lag one model of Rydberg and Shepard to include an arbitrary number of lags for both the past counts and past conditional means and named it the Autoregressive Conditional Poisson model with lags $p$ and $q$ (ACP ($p$, $q$)). The formulation of this model resembles that of the generalized conditional heteroscedastic (GARCH) model of Bollerslev (1986), but unlike the GARCH formulation that models the conditional variance of the process, the ACP models the conditional mean. Heinen, however, derived the properties of his model only for the ACP (1, 1) case. The general case was investigated by Ghahramani and Thavaneswaran (2009) who referred to the Heinen paper as the origin of the ACP model. Independently, Ferland, Latour, and Oraichi (2006) proposed what was termed the Integer GARCH (INGARCH) process, which is essentially the same as the ACP model of Heinen. The INGARCH (or ACP) model, however, does not accommodate zero-inflation, and Zhu (2012a) proposed a zero-inflated INGARCH formulation to facilitate the modeling of count data with zero counts that cannot be fitted well by the regular INGARCH model. The zero-inflation probability in Zhu's model is constant over time. This is a drawback in situations where the relative number of zero counts tends to vary seasonally or influenced by an exogenous variable. The proposed Time-Varying Zero-inflated Poisson INGARCH model (TVZIP-INGARCH) was developed to address this shortcoming. Before introducing the proposed model, we provide a brief overview of literature in the following paragraphs. For brevity, this overview will not cover the extensive literature on count data time series and will be limited to what is directly related the proposed model.

Integer-valued Autoregressive (INAR) Poisson models were introduced by McKenzie (1985) as well as by Al-Osh and Alzaid (1987) for analyzing equidispersed count data with serial correlation. Quddus (2008) conducted an empirical study using an INAR model to analyze traffic accidents in Great Brittan and compared the performance with the results from fitting a real-valued Autoregressive Integrated Moving Average (ARIMA) model. The study found that the INAR Poisson model performed better when the counts are relatively low. Another formulation developed for analyzing count data time series is the Generalized Linear Autoregressive Moving Average (GLARMA) model (Davis, Dunsmuir, and Streett, 2003). In this model the logarithm of the intensity of the Poisson process is assumed to depend on the past counts as well as current and past values of explanatory variables, with the noise process rewritten as an ARMA model. In contrast, the ACP model assumes that the counts are generated via a Poisson distribution with its conditional mean obeying a dynamic process akin to that of the classical Generalized Autoregressive Conditional Heteroscedastic (GARCH) model. It also has a connection to the Autoregressive Conditional Duration (ACD) model (Engle and Russell, 1998) which models duration between events. In the ACD model, the conditional mean duration between events dependents on past durations as well as past conditional means. The link between the two models can be made by taking inverse of the conditional mean duration modeled by the ACD formulation as the intensity of the Poisson process in the ACP model. Because the unconditional variance of the count variable in the ACP model is higher than the unconditional expectation, this class of models accommodates over dispersion while at the same time accounting for autocorrelation. As mentioned previously, the ACP model is exactly the same as the model proposed by Ferland *et al.* (2006). When discussing published literature on this topic, we will refer to these models as ACP or INGARCH, interchangeably, based on what the author(s) of the cited works used in referring to the model. When not discussing a particular publication, it will be referred to as the INGARCH model.

The INGARCH model of order ($p$, $q$) is defined as follows:

$$X_t | \mathbb{F}_{t-1} \sim P(\lambda_t); \forall t \in \mathbb{Z}$$
$$\lambda_t = \alpha_0 + \sum_{i=1}^{p} \alpha_i X_{t-i} + \sum_{j=1}^{q} \beta_j \lambda_{t-j}, \quad (1.1)$$

where $\{X_t : t \in \mathbb{Z}\}$ is the count process, $\lambda_t$ defines the conditional mean of $X_t$ given the sigma-field $\mathbb{F}_{t-1}$ generated by $X_l$ and $\lambda_l$ values for $l<t$. The model is defined with the parameters constraints $\alpha_0 > 0, \alpha_i \geq 0, \beta_j \geq 0, i = 1, ..., p, j = 1, ..., q, p \geq 1, q \geq 0,$ and $0 \leq \sum_{i=1}^{p} \alpha_i + \sum_{j=1}^{q} \beta_j < 1.$ In addition to presenting the model, Heinen (2003) also derived the stationarity conditions, covariance functions, and addressed the problem of maximum likelihood estimation (MLE) of the parameters.

Having introduced the INGARCH model, we proceed with a summary of additional work related to this formulation. To address the scarcity of literature available for the general class of INGARCH ($p$, $q$) models, Weiß (2009) extended the previous results and derived a set of Yule-Walker type equations for the autocorrelation function for the general INGARCH case. Several important theoretical contributions for both the linear and non-linear Poisson autoregressive models were made by Fokianos, Rahbek, and Tjøstheim (2009). For the model where the conditional mean is a linear function of the past conditional mean and past count, they proved that the maximum likelihood estimates are asymptotically normal. Note that the model they considered is INGARCH (1, 1), but the authors refrain from calling it as such and labeled it as Poisson autoregression because of their stated belief that the GARCH moniker should be reserved for formulations that model variance. In the non-linear case where the conditional mean is a non-linear function of its past values and past counts, the authors established that the intensities of the Poisson distribution at each time point form a geometrically ergodic Markov chain under some general assumptions. Fokianos and Tjøstheim (2011) further extended these results and showed that in the log-linear Poisson autoregression case, the maximum likelihood estimates are asymptotically normal, and that covariance matrix of the parameter estimates are consistent. Additional work on the INGARCH models includes Zhu and Wang (2011) that considered testing the parameters of a Poisson autoregressive model. As pointed out by several authors (Weiß, 2009, and Zhu, 2012b), the INGARCH formulations are frequently utilized in modeling the over-dispersion and serial dependency inherent to count data. Zhu (2011) discussed the modelling of integer valued time series with over dispersion and handling potential extreme observations and Zhu (2012b) generalized the Poisson INGARCH process to handle both over dispersion and under dispersion cases. Further, the author provided real-life applications of the proposed model. Fokianos and Fried (2012) applied the log-linear Poisson autoregressive model, with an additional term, to two real-life data sets with a goal of detecting unusual events, or what they termed "interventions." The additional term they included in the model is the key to identifying intervention types such as level shifts and transient shifts when the time of intervention is unknown. This work complements the work presented in Fokianos and Fried (2010) for the INGARCH models.

Negative binomial (NB), Generalized Poisson (GP), and Double Poisson (DP) are well known discrete distributions that can also be used as an alternative to the Poisson process (Zhu, 2012c). A negative binomial INGARCH model (NB-INGARCH), which is an alternative to the Poisson INGARCH model, was proposed and the stationary conditions and the autocorrelation function of the process were obtained by Ye, Garcia, Pourahmadi, and Lord (2012). These authors also allowed the negative binomial INGRACH model to incorporate covariates, so that the relationship between a time series of counts and correlated external factors could be properly modeled. Zhu (2012a) while introducing the zero-inflated Poisson, also presented a zero-inflated negative Binomial integer-valued GARCH model and showed how the EM algorithm can be used to estimate the parameters. The underlying processes in Zhu's models are based on either zero-inflated Poisson or zero-inflated negative binomial, but they do not allow such zero-inflation to vary over time or influenced by any external factor.

Some empirical time series count data with large number of zero counts display strong cyclical behavior or seasonality with respect to the observed zero values. Ignoring this time varying property of the zero-inflation parameter decreases the predictive performance of the model. Recognizing this, Yang (2012), discussed the importance of modeling zero-inflation as a time varying function. In the above article, it was assumed that both the zero-inflation and the intensity parameter are driven by the linear combination of past observations of exogenous variables and connects them to the conditional mean of the count via a log-link function. A recently introduced approach to modeling time-varying zero-inflation and the intensity parameter is the adaptive log-linear zero-inflated generalized Poisson model, denoted by the abbreviation INGARCHX (ALG), proposed by Xu, Y. Chen, C. Chen, and Lin (2020). Here X denotes a model with exogenous inputs. The authors assumed that the counts are from a zero-inflated generalized Poisson distribution, with the logarithm of the intensity propagated through a GARCH type model augmented with additional terms of exogenous variables and associated coefficients. All model parameters are assumed to be time depended. If this time dependance is removed, then the model becomes the log-linear Poisson autoregressive formulation proposed by Fokianos and Tjøstheim (2011). In applying to monthly crime data from New South Wales, Australia, the authors assumed constant coefficients over a local interval at any given time point *t* (with the interval adaptively selected from among a prechosen set of nested intervals) and estimated the parameters separately using data from each such interval. In their approach, the shifting of the intervals allows for parameters to change from one interval to another and thus allowing for the zero-inflation as well as other model parameters to vary with time.

We propose a different approach based on a generalization of the model proposed by Zhu (2012a). In our formulation, it is the zero-inflation probability, rather than the mean of the Poisson process, that is allowed to be governed by exogenous variables. We also assume that the INGARCH model parameters remain constant over time, allowing for a more parsimonious model that is relatively easier to estimate. While the model proposed by Xu *et al*. (2020) is more flexible, the approach we propose is presented as a simpler and practical alternative, albeit a less sophisticated one. The proposed model can also accommodate the case where the zero-inflation probability is driven by a deterministic function of time, such as a sinusoidal wave. In addition, the conditional mean of the Poisson process is assumed to vary dynamically through a GARCH type model. Thus, the INGARCH part of the proposed model can be viewed as observation driven, in the sense that recursive substitutions can be employed to show that the current mean of the process conditional on the past is a linear function of past observations and conditional means. Also note that we model the conditional mean rather than the log of the conditional mean as is done in Xu *et al.* (2020).

The remainder of this paper is organized as follows: In Section 2, the Time Varying Zero Inflated Poisson INGARCH model (TVZIP-INGARCH) is introduced. Two cases, namely a deterministic cyclically varying zero-inflation component and a model with the zero-inflation parameter driven by an exogenous set of stochastic variables is discussed. Parameter estimation procedures are presented in Section 3, which is followed by a simulation study presented in Section 4. Results of fitting the proposed model to empirical data are presented in Section 5. A discussion and conclusions are presented in Section 6.

## 2. THE TIME VARYING ZERO-INFLATED INGARCH MODEL

As Zhu (2012a) stated, the probability mass function (pmf) of a zero-inflated Poisson model with parameter vector $(\lambda, \omega)$, with $X$ representing the count, can be written in the following form:

$$P(X = k) = \omega \delta_{k,0} + (1-\omega) \frac{\lambda^k e^{-\lambda}}{k!}, \ k = 0, 1, 2, ..., \text{where } 0 < \omega < 1 \text{ and}$$

$$\delta_{k,0} = \begin{cases} 1; k = 0 \\ 0; k \neq 0. \end{cases}$$

Further, Zhu (2012a) presented the mean and the variance of the distribution as follows:

$$E(X) = \lambda(1-\omega) \text{ and } Var(X) = \lambda(1-\omega)(1+\lambda\omega) > E(X) \text{ for } 0 < \omega < 1.$$

Moving on to define the time-varying zero-inflated INGARCH model, assume that $\{X_t : t \in \mathbb{Z}\}$ is a discrete time series of count data, and $\mathbb{F}_{t-1}$ is the sigma field generated by $\{X_l : l \leq t-1\}$. The conditional distribution of $X_t$ given $\mathbb{F}_{t-1}$ is assumed to be a zero-inflated Poisson (*ZIP*) with parameter vector $(\lambda_t, \omega_t)$. Then, $X_t \mid \mathbb{F}_{t-1} \sim ZIP(\lambda_t, \omega_t)$ where,

$$P(X_t = k \mid \mathbb{F}_{t-1}) = \omega_t \delta_{k,0} + (1 - \omega_t) \frac{\lambda_t^k e^{-\lambda_t}}{k!}. \qquad (2.1)$$

The dynamic propagation of the conditional mean of the Poisson process is defined by

$$\lambda_t = \alpha_0 + \sum_{i=1}^{p} \alpha_i X_{t-i} + \sum_{j=1}^{q} \beta_j \lambda_{t-j},$$

where $\alpha_0 > 0, \alpha_i \geq 0, \beta_j \geq 0, i = 1, 2, 3, \ldots p, j = 1, 2, 3, \ldots, q, p \geq 1, q \geq 1,$ and $t \in \mathbb{Z}$.

Furthermore, $\omega_t = g(\mathbf{V}_t, \underline{\Gamma}) \in (0,1) \ \forall \ t \in \mathbb{Z}$, is a function of variables, propagating over time, which is used to model the time varying zero-inflation. Note that elements of the vector $\mathbf{V}_t$ may consist of stochastic exogenous variables that vary with time, or it may be a scaler equal to time $t$. In addition, $\underline{\Gamma}$ denotes a vector of parameters. It is assumed that $0 < \omega_t < 1$ for all $t \in \mathbb{Z}$. Note that Fokianos *et al.* (2009) defined their linear Poisson autoregressive model for $t \in \mathbb{N}$ instead of $t \in \mathbb{Z}$, with constant initial conditions. The model in (2.1) can also be defined in a similar manner. It is this alternative formulation that was used in our simulation study.

The above model is denoted by TVZIP-INGARCH ($p$, $q$). If $p > 0$ and $q = 0$, then the model becomes a TVZIP-INARCH model with order $p$, denoted by TVZIP-INARCH ($p$). The conditional mean and conditional variance of $X_t$ given $\mathbb{F}_{t-1}$ are specified by the following equations:

$$E(X_t \mid \mathbb{F}_{t-1}) = (1 - \omega_t)\lambda_t \text{ and } Var(X_t \mid \mathbb{F}_{t-1}) = (1 - \omega_t)\lambda_t(1 + \omega_t \lambda_t). \qquad (2.2)$$

See Appendix A.1, for the derivation of the conditional mean and conditional variance.

The conditional variance to conditional mean ratio, or the dispersion ratio, of ZIP distribution is:

$$\frac{Var(X_t \mid \mathbb{F}_{t-1})}{E(X_t \mid \mathbb{F}_{t-1})} = \frac{(1-\omega_t)\lambda_t(1+\omega_t\lambda_t)}{(1-\omega_t)\lambda_t} > (1+\lambda_t\omega_t). \tag{2.3}$$

The result in (2.3) indicates that TVZIP-INGARCH ($p$, $q$) can be used to model integer valued time series with over dispersion if the values of $\omega_t$ and $\lambda_t$ are uniformly bounded below by positive constants.

## 2.1. CASE 1: ZERO-INFLATION DRIVEN BY A DETERMINISTIC FUNCTION OF TIME.

In Case 1, it is assumed that the zero-inflation function $\omega_t = g(\mathbf{V}_t, \underline{\Gamma})$ is such $\mathbf{V}_t$ is a scaler equal to $t$. For example, we may assume the function $g$ to be defined as follows:

$$\omega_t = g(\mathbf{V}_t, \underline{\Gamma}) = A\sin\left(\frac{2\pi}{s}t\right) + B\cos\left(\frac{2\pi}{s}t\right) + C \tag{2.4}$$

where $s$ is the seasonal length, and $\underline{\Gamma} = \begin{pmatrix} A \\ B \\ C \end{pmatrix}$. Here $A, B, C \in \mathbb{R}$.

As mentioned above, the time-varying zero-inflation function $\omega_t = g(\mathbf{V}_t, \underline{\Gamma})$ should always be bounded between zero and one. The range of values for $A, B$, and $C$ in (2.4) that are needed to satisfy the above criterion are derived in Appendix B. Note that herein a simple example is used, where the function $g$ consists of a sine function and a cosine function of equal period, but $g$ could be any other function of time that, with proper selection of parameters, can be bounded between zero and one.

## 2.2. CASE 2: ZERO-INFLATED FUNCTION DRIVEN BY AN EXOGENOUS VARIABLES

The proposed model also accommodates the case where the zero-inflation probability is determined by one or more exogenous variables. In this case $g(\mathbf{V}_t, \underline{\Gamma})$ is considered a function, with the interval (0, 1) as its domain, of the vector of exogenous variables $\mathbf{V}_t$. One example of $g$ is the logistic function. Note that $\mathbf{V}_t$ can be a scaler seasonal autoregressive time series, a vector seasonal time series, or a scaler or vector time series that varies non-seasonally. For illustrative purposes, consider the case where $\mathbf{V}_t$ is a scaler purely seasonal autoregressive time series, denoted by $V_t$, with period $s$, $g$ the logistic function, and $\varepsilon_t$ is a white noise error term. Then we can write,

$$V_t = \eta V_{t-s} + \varepsilon_t, \text{where } \varepsilon_t \sim WN(0,1);$$

$$\omega_t = g(V_t, \underline{\Gamma}) = \frac{1}{1 + e^{-(\delta_0 + \delta_1 V_t)}}, \quad (2.5)$$

with $\underline{\Gamma} = \begin{pmatrix} \delta_0 \\ \delta_1 \end{pmatrix}$. Here $A, B,$ and $C \in \mathbb{R}$.

### 3. ESTIMATION PROCEDURE

The use of both the Expectation Maximization (EM) algorithm and Maximum Likelihood (ML) method to estimate the model parameters were developed for the general TVZIP-INGARCH ($p$, $q$) case. For brevity, only the TVZIP-INGARCH (1, 1) process is discussed below, but the procedure for the general case follows in a similar manner, even though the computations would be more complex.

### 3.1. EXPECTATION MAXIMIZATION ESTIMATION FOR THE TVZIP-INGARCH (1, 1) PROCESS

Let $X_1, X_2, ..., X_N$ be generated according to the model (2.1). There are two types of zeros generated by this model. They are the zeroes arising from the Poisson distribution with intensity parameter $\lambda_t$ and the zeroes generated by a Bernoulli process with the probability of obtaining a zero specified by the zero-inflation parameter. Therefore, a given observation can be hypothetically categorized as arising out of a Bernoulli process or as an observation from the Poisson distribution. Let us define $\{Z_t \mid t \in \mathbb{Z}\}$ to be a Bernoulli random variable such that $Z_t = 1$ if $X_t$ is a generated from the Bernoulli process and $Z_t = 0$ if it is generated by the Poisson distribution. Then,

$$Z_t \sim \text{Bernoulli}(\omega_t) \text{ with } P(Z_t = 1) = \omega_t \text{ and } P(Z_t = 0) = (1 - \omega_t).$$

Also, let $Z = (Z_1, Z_2, ..., Z_N)$, $\Theta = (\alpha_0, \alpha_1, ..., \alpha_p, \beta_1, \beta_2, ..., \beta_q)^T = (\theta_0, \theta_1, ..., \theta_{p+q})$ and $\omega_t = g(\mathbf{V}_t, \underline{\Gamma})$. Note that $\Gamma = (\gamma_0, \gamma_1, ..., \gamma_r)$, where $r$ is the dimension of the vector $\mathbf{V}_t$. For notational simplicity, we define the composite parameter vector $\Phi = (\Gamma^T, \Theta^T)^T = (\phi_1, \phi_2, ......, \phi_{r+p+q+2}) \subseteq \mathbb{R}^{r+p+q+2}$, with the original parameters renamed as $\phi_k$, $k = 1, 2, ..., r+p+q+2$. This simplified notation is used in situations where generic statements are made without reference to a specific portion of (2.1).

Paralleling the derivations in Zhu (2012a), the conditional log likelihood can be written as (see Appendix A.2 for details),

$$l(\Phi) = \sum_{t=p+1}^{N} \left\{ Z_t \log(\omega_t) + (1 - Z_t)\left[\log(1 - \omega_t) + X_t \log(\lambda_t) - \lambda_t - \log(X_t!)\right]\right\}. \quad (3.1)$$

The first derivatives of the conditional log likelihood function (3.1) with respect to $\Gamma = (\gamma_0, \gamma_1, ..., \gamma_r)$ and $\Theta = (\theta_0, \theta_1, ..., \theta_{p+q})$ are as follows:

$$\frac{dl(\Phi)}{d\gamma_i} = \frac{dl(\Phi)}{d\omega_t} \frac{d\omega_t}{d\gamma_i} = \sum_{t=p+1}^{N} \left\{ \frac{Z_t}{\omega_t} - \frac{(1 - Z_t)}{(1 - \omega_t)} \right\} \frac{d\omega_t}{d\gamma_i}, \quad i = 0, 1, ..., r, \quad (3.2)$$

$$\frac{dl(\Phi)}{d\theta_j} = \frac{dl(\Phi)}{d\lambda_t} \frac{d\lambda_t}{d\theta_j} = \sum_{t=p+1}^{N} (1 - Z_t)\left\{ \frac{X_t}{\lambda_t} - 1 \right\} \frac{d\lambda_t}{d\theta_j}, \quad j = 0, 1, ..., p+q. \quad (3.3)$$

Finally, by combining (3.2) and (3.3) the first derivative of the conditional log likelihood function with respect to $\Phi$ is given by:

$$\frac{dl(\Phi)}{d\phi_k} = \frac{dl(\Phi)}{d\omega_t}\frac{d\omega_t}{d\phi_k} + \frac{dl(\Phi)}{d\lambda_t}\frac{d\lambda_t}{d\phi_k}; \phi_k \in \Phi = (\Gamma, \Theta)$$

and $\quad \dfrac{d\omega_t}{d\phi_k} = 0$, if $\phi_k \notin \Gamma$ and $\dfrac{d\lambda_t}{d\phi_k} = 0$ if $\phi_k \notin \Theta$. (3.4)

The two-step (E step and M step) Expectation Maximization algorithm is now used to estimate the parameter vector $\Phi = (\Gamma^T, \Theta^T)^T$. Let $\tau_t = E(Z_t \mid X_t, \Phi)$ and we replace $Z_t$ by $\hat{Z}_t = \tau_t$ and define $Z = (Z_1, Z_2, \ldots, Z_N)^T$. Following this replacement of Z in the conditional log likelihood function, $l(\Phi, \hat{Z})$ is maximized.

**E Step:** Determine $\tau_t$ using the equation

$$\tau_t = \begin{cases} \dfrac{\omega_t}{\omega_t + (1-\omega_t)e^{-\lambda_t}} & : X_t = 0 \\ 0 & : X_t > 0. \end{cases}$$

**M Step:** After $Z_t$ is replaced by its estimate, we proceed to maximize $l(\Phi, \hat{Z})$. First set

$$\dfrac{dl(\Phi)}{d\phi_k} = 0, \text{ for } k = 1, 2, \ldots, r+p+q+2.$$

If, $\hat{\Phi}$, the solution to the system of equations in (3.4) exists, then $S(\hat{\Phi}) = \underline{0}$, where $S(\Phi)$ is the Fisher's score matrix, and $\hat{\Phi}$ is the vector that minimizes the log likelihood, thus providing us with the estimate of the parameter vector $\Phi = (\Gamma^T, \Theta^T)^T$.

Since a closed form solution does not exist, we require an iterative procedure to find the estimates. Let us consider the first order Taylor expansion of $S(\tilde{\Phi})$ evaluated at the value $\tilde{\Phi}$ around the initial parameter values $\Phi_0$, yielding. $S(\tilde{\Phi}) \approx S(\Phi_0) + \dfrac{dS(\Phi)}{d\phi}(\tilde{\Phi} - \Phi_0)$. We also let the matrix of the second derivatives of the log likelihood function to be defined as $H(\Phi) = \dfrac{d^2 l(\Phi)}{d\Phi d\Phi^T} = \dfrac{dS(\Phi)}{d\Phi}$.

From the above, we obtain the first order approximation $\tilde{\Phi} = \Phi_0 - H^{-1}(\Phi_0)S(\Phi_0)$, and this result provides the standard Newton-Raphson algorithm. For an appropriately chosen initial value $\Phi_0$, the above Newton Raphson algorithm can be used to obtain a sequence of improved estimates recursively. The improved estimates at $i^{th}$ iteration are updated as the initial values for the next iteration as follows:

$$\hat{\Phi}^{(i+1)} = \hat{\Phi}^{(i)} - H^{-1}(\hat{\Phi}^{(i)})S(\hat{\Phi}^{(i)}).$$

This Process is repeated until the differences between successive estimates are sufficiently close to zero. In our study, convergence of the EM procedure was determined by using the criterion:

$$\left| \left(\hat{\phi}_j^{(i+1)} - \hat{\phi}_j^{(i)}\right) \Big/ \left(\hat{\phi}_j^{(i)}\right) \right| \leq 10^{-6}.$$

## 3.2. MAXIMUM LIKELIHOOD ESTIMATION FOR THE TVZIP-INGARCH (1, 1) PROCESS

The conditional likelihood function $L(\Phi)$ of the TVZIP-INGARCH model (2.1) is,

$$L(\Phi) = \prod_{X_t=0}\left[\omega_t + (1-\omega_t)e^{-\lambda_t}\right]\prod_{X_t>0}\left[(1-\omega_t)\frac{\lambda_t^{X_t}e^{-\lambda_t}}{X_t!}\right]. \quad (3.5)$$

The conditional log likelihood function, $l(\Phi)$ obtained from (3.5) is given by

$$l(\Phi) = \sum_{X_t=0}\log\left|\omega_t + (1-\omega_t)e^{-\lambda_t}\right| + \sum_{X_t>0}\left[\log(1-\omega_t) + X_t\log(\lambda_t) - \lambda_t - \log(X_t!)\right]. \quad (3.6)$$

Let $P_{0,t} = \omega_t + (1-\omega_t)e^{-\lambda_t}$ and $I(X_t = 0) = x_{0,t}$. Then,

$$\frac{dl(\Phi)}{d\omega_t} = \sum_{t=p+1}^{N}\left[\frac{x_{0,t}(1-e^{-\lambda_t})}{P_{0,t}} - \frac{(1-x_{0,t})}{\lambda_t}\right], \quad (3.7)$$

and

$$\frac{dl(\Phi)}{d\lambda_t} = \sum_{t=p+1}^{N}\left[\frac{x_{0,t}\omega_t}{P_{0,t}} + \frac{(X_t - \lambda_t)}{\lambda_t}\right]. \quad (3.8)$$

The first derivatives of the conditional log likelihood function (3.6) are as follows,

$$\frac{dl(\Phi)}{d\gamma_i} = \frac{dl(\Phi)}{d\omega_t}\frac{d\omega_t}{d\gamma_i} = \sum_{t=p+1}^{N}\left[\frac{x_{0,t}(1-e^{-\lambda_t})}{P_{0,t}} - \frac{(1-x_{0,t})}{\lambda_t}\right]\frac{d\omega_t}{d\gamma_i}, i=0,1,...,r, \quad (3.9)$$

$$\frac{dl(\Phi)}{d\theta_j} = \frac{dl(\Phi)}{d\lambda_t}\frac{d\lambda_t}{d\theta_j} = \sum_{t=p+1}^{N}\left[\frac{x_{0,t}\omega_t}{P_{0,t}} + \frac{(X_t - \lambda_t)}{\lambda_t}\right]\frac{d\lambda_t}{d\theta_j}, j=0,1,...,p+q, \quad (3.10)$$

$$\frac{dl(\Phi)}{d\phi_k} = \frac{dl(\Phi)}{d\omega_t}\frac{d\omega_t}{d\phi_k} + \frac{dl(\Phi)}{d\lambda_t}\frac{d\lambda_t}{d\phi_k}, \phi_k \in \Phi = (\Gamma, \Theta),$$

$$\frac{d\omega_t}{d\phi_k} = 0, \text{ if } \phi_k \notin \Gamma \;\; \phi_k \notin \Gamma \text{ and } \frac{d\lambda_t}{d\phi_k} = 0, \text{ if } \phi_k \notin \Theta \quad (3.11)$$

We can use Newton-Raphson (NR) iterative procedure to obtain the maximum likelihood estimated for the equation (3.5) by setting $\frac{dl(\Phi)}{d\phi_k} = 0$ for all $k$. With a reasonable initial starting value $\hat{\Phi}^{(0)}$, the $i^{\text{th}}$ iteration is calculated using $\hat{\Phi}^{(i+1)} = \hat{\Phi}^{(i)} - H^{-1}(\hat{\Phi}^{(i)})S(\hat{\Phi}^{(i)})$, where $S(\hat{\Phi}) = \frac{dl(\Phi)}{d\phi_k}\Big|\hat{\Phi}$ and $H(\hat{\Phi}) = \frac{dl(\Phi)}{d\phi_k d\phi_k^T}\Big|\hat{\Phi}$.

We stop the algorithm once pre specified convergence criteria is satisfied.

## 4. SIMULATION STUDY

We investigated the finite sample performance of estimators using a simulation study. The *poissrnd* function of MATLAB software was employed to generate the relevant data, based on recursively computed conditional mean. In order to initiate the recurve process, the conditional mean at time $t=0$ and count data at times $t \leq 0$ were set to zero (i.e., $\lambda_0 = 0$ and $X_l = 0$, for $l \leq 0$). For time periods $t \geq 1$, $X_t$ was generated as follows. For each $t \in \mathbb{N}$ let $U_t$ be a random variable generated from a uniform $(0,1)$ distribution and let $\omega_t$ be the zero-inflated probability at time $t$. Then, $X_t$ was set to zero if $U_t \leq \omega_t$, Otherwise $X_t$ was generated from the Poisson distribution with intensity parameter $\lambda_t$, where $\lambda_t$ was updated recursively using Equation (2.1). The process was repeated until the complete time series of length $N$ was generated. Note that this is the same procedure Zhu (2012a) employed to generate data for his simulation study. Lengths of the time series studied were set to $N = 120$ and $N = 360$, and thousand ($m = 1000$) simulations runs were carried out for each parameter and sample size combination. We carried out two separate sets of simulation studies based on the two types of zero-inflation function introduced in Section 2. The profile log likelihood function given in Equation (3.6) was maximized using the constrained nonlinear optimization function *fmincon* in MATLAB. The zero-inflation probability $(\omega_t = g(\mathbf{V}_t, \Gamma))$ was allowed to vary cyclically as a deterministic function of time or to be driven by an exogenous variable. Following Zhu (2012a), the Mean Absolute Deviation Error (MADE) was utilized as the evaluation criterion. The MADE is defined as, $\frac{1}{m}\sum_{j=1}^{m}|\hat{\phi}_j - \phi|$ where $m$ is the number of replications and $\phi \in \Phi = (\Gamma^T, \Theta^T)^T$ is the true value while $\hat{\phi}_j$ is the estimated value of $\phi$ at $j^{\text{th}}$ replication run. Simulation results are reported in Tables 1 through Table 12.

## 4.1. SIMULATION RESULTS FOR CASE 1: DETERMINISTIC SINUSOIDAL ZERO-INFLATION FUNCTION

In this portion of the simulation study, the sinusoidal zero-inflated function $\omega_t = g(\mathbf{V}_t, \Gamma)$ expressed in Equation (2.4) was used to generate cyclically varying zero-inflation probabilities between zero and one. We set the following constraints to the parameters in the vector $\Gamma = \begin{pmatrix} A \\ B \\ C \end{pmatrix}$:

$C = \sqrt{A^2 + B^2} + \delta$, $\sqrt{A^2 + B^2} \leq \frac{1}{2} - \delta$, where $|A| \leq \frac{1}{2} - \delta$, $|B| \leq \frac{1}{2} - \delta$, and $\delta$ a constant such that $\delta \in \left(0, \frac{1}{2}\right)$. Note that the above constraints, with $\delta = 0.0001$, were applied in our simulation study in order to bound the zero-inflation probabilities between 0 and 1. A very small value for $\delta$ was selected to allow wider bounds for $\sqrt{A^2 + B^2}$, $|A|$, and $|B|$.

Tables 1 through 3 provide the simulation results for the MLE estimation technique, while Tables 4 through 6 provide simulation results for the case where estimates were obtained using the EM algorithm. The frequency of the sinusoidal wave was set at $s = 12$, mimicking a 12-month cycle present in monthly data. The parameter vector for the simulation study was expressed as $\Phi = (A, B, \alpha_0, \alpha_1, \alpha_2, \beta_1)$, where $A$ and $B$ are the parameters in the sinusoidal model while $(\alpha_0, \alpha_1)$, $(\alpha_0, \alpha_1, \alpha_2)$, and $(\alpha_0, \alpha_1, \beta_1)$ are the parameter combinations in TVZIP-INARCH (1), TVZIP-INARCH (2), TVZIP-INGARCH (1, 1) models, respectively. The parameter combination of $\Gamma = (A, B)^T$ was set at (0.10, 0.10), (0.25, -0.20), and (-0.35, -0.30) and represent minimal to minimal, minimal to moderate, and minimal to maximum zero-inflation ranges.

The following models were considered:

**(A)** TVZIP-INARCH (1) models: $\Phi = (A, B, \alpha_0, \alpha_1)$

  A1.   (0.10, 0.10, 1.00, 0.40)
  A2.   (-0.25, -0.25, 2.00, 0.50)
  A3.   (-0.35, -0.30, 1.00, 0.70)

**(B)** TVZIP-INARCH (2) models: $\Phi = (A, B, \alpha_0, \alpha_1, \alpha_2)$

  B1.   (0.10, 0.10, 1.00, 0.20, 0.20)
  B2.   (-0.25, -0.25, 2.00, 0.30, 0.20)
  B3.   (-0.35, -0.30, 1.00, 0.40, 0.30)

**(C)** TVZIP-INGARCH (1,1) models: $\Phi = (A, B, \alpha_0, \alpha_1, \beta_1)$

  C1.   (0.10, 0.10, 1.00, 0.20, 0.20)

C2.   (-0.25, -0.25, 2.00, 0.30, 0.20)

C3.   (-0.35, -0.30, 1.00, 0.40, 0.30)

Note that the two estimations procedures were run on identical simulation samples for each model, parameter, and sample size combinations and hence variations due to sampling error will not be seen when comparing across estimation methods.

Table 1: Means of MLE estimates and MADE (within parentheses), for TVZIP-INARCH (1) models where zero-inflation is driven by a sinusoidal function.

| Model | N | A | B | $\alpha_0$ | $\alpha_1$ |
|---|---|---|---|---|---|
| | **True values** | **0.10** | **0.10** | **1.00** | **0.40** |
| A1 | 120 | 0.0893 (0.0561) | 0.0845 (0.0566) | 1.0472 (0.1433) | 0.3712 (0.0903) |
| A1 | 360 | 0.0986 (0.0321) | 0.0951 (0.0299) | 1.0172 (0.0802) | 0.3917 (0.0479) |
| | **True Values** | **-0.25** | **-0.25** | **2.00** | **0.50** |
| A2 | 120 | -0.2488 (0.0401) | -0.2475 (0.0399) | 2.0467 (0.2210) | 0.4800 (0.0873) |
| A2 | 360 | -0.2504 (0.0225) | -0.2478 (0.0223) | 2.0203 (0.1311) | 0.4925 (0.0484) |
| | **True Values** | **-0.35** | **-0.30** | **1.00** | **0.70** |
| A3 | 120 | -0.3468 (0.0457) | -0.2959 (0.0482) | 1.0369 (0.1621) | 0.6675 (0.1163) |
| A3 | 360 | -0.3508 (0.0250) | -0.2970 (0.0263) | 1.0178 (0.0963) | 0.6875 (0.0629) |

Table 2: Means of MLE estimates and MADE (within parentheses), for TVZIP-INARCH (2) models where zero-inflation is driven by a sinusoidal function.

| Model | N | A | B | $\alpha_0$ | $\alpha_1$ | $\alpha_2$ |
|---|---|---|---|---|---|---|
| | **True values** | **0.10** | **0.10** | **1.00** | **0.20** | **0.20** |
| B1 | 120 | 0.0872 (0.0559) | 0.0838 (0.0538) | 1.0529 (0.1602) | 0.1877 (0.0873) | 0.1765 (0.0864) |
| B1 | 360 | 0.0955 (0.0317) | 0.0971 (0.0290) | 1.0239 (0.0945) | 0.1943 (0.0496) | 0.1906 (0.0503) |
| | **True Values** | **-0.25** | **-0.25** | **2.00** | **0.30** | **0.20** |
| B2 | 120 | -0.2485 (0.0430) | -0.2476 (0.0395) | 2.0524 (0.2563) | 0.2842 (0.0952) | 0.1906 (0.0931) |
| B2 | 360 | -0.2514 (0.0234) | -0.2470 (0.0224) | 2.0254 (0.1478) | 0.2989 (0.0500) | 0.1901 (0.0536) |
| | **True Values** | **-0.35** | **-0.30** | **1.00** | **0.40** | **0.30** |
| B3 | 120 | -0.3478 (0.0455) | -0.2941 (0.0486) | 1.0357 (0.1712) | 0.3839 (0.1305) | 0.2781 (0.1324) |
| B3 | 360 | -0.3510 (0.0282) | -0.2974 (0.0273) | 1.0149 (0.1004) | 0.3975 (0.0709) | 0.2914 (0.0762) |

Table 3: Means of MLE estimates and MADE (within parentheses), for TVZIP-INGARCH (1, 1) models where zero-inflation is driven by a sinusoidal function.

| Model | $N$ | $A$ | $B$ | $\alpha_0$ | $\alpha_1$ | $\beta_1$ |
|---|---|---|---|---|---|---|
| | **True values** | **0.10** | **0.10** | **1.00** | **0.20** | **0.20** |
| **C1** | 120 | 0.0939 (0.0521) | 0.0882 (0.0541) | 0.9420 (0.2205) | 0.2226 (0.0789) | 0.2078 (0.1399) |
| | 360 | 0.1057 (0.0300) | 0.0983 (0.0295) | 0.9511 (0.1544) | 0.2295 (0.0554) | 0.1977 (0.1162) |
| | **True Values** | **-0.25** | **-0.25** | **2.00** | **0.30** | **0.20** |
| **C2** | 120 | -0.2522 (0.0396) | -0.2472 (0.0402) | 1.7725 (0.3498) | 0.3609 (0.0979) | 0.1929 (0.1178) |
| | 360 | -0.2516 (0.0231) | -0.2500 (0.0221) | 1.8088 (0.2330) | 0.3843 (0.0884) | 0.1540 (0.0859) |
| | **True Values** | **-0.35** | **-0.30** | **1.00** | **0.40** | **0.30** |
| **C3** | 120 | -0.3589 (0.0414) | -0.2937 (0.0456) | 0.9463 (0.2131) | 0.4666 (0.1294) | 0.2327 (0.1689) |
| | 360 | -0.3591 (0.0259) | -0.2984 (0.0265) | 0.9426 (0.1442) | 0.4958 (0.1088) | 0.2022 (0.1372) |

Table 4: Means of EM estimates and MADE (within parentheses), for TVZIP-INARCH (1) models where zero-inflation is driven by a sinusoidal function.

| Model | N | A | B | $\alpha_0$ | $\alpha_1$ |
|---|---|---|---|---|---|
| | **True values** | **0.10** | **0.10** | **1.00** | **0.40** |
| **A1** | 120 | 0.0898 (0.0557) | 0.0850 (0.0561) | 1.0472 (0.1433) | 0.3712 (0.0903) |
| | 360 | 0.0986 (0.0321) | 0.0951 (0.0299) | 1.0172 (0.0802) | 0.3917 (0.0479) |
| | **True Values** | **-0.25** | **-0.25** | **2.00** | **0.50** |
| **A2** | 120 | -0.2487 (0.0401) | -0.2476 (0.0398) | 2.0467 (0.2210) | 0.4800 (0.0873) |
| | 360 | -0.2504 (0.0225) | -0.2478 (0.0222) | 2.0203 (0.1311) | 0.4925 (0.0484) |
| | **True Values** | **-0.35** | **-0.30** | **1.00** | **0.70** |
| **A3** | 120 | -0.3468 (0.0457) | -0.2959 (0.0482) | 1.0369 (0.1621) | 0.6675 (0.1164) |
| | 360 | -0.3508 (0.0250) | -0.2970 (0.0263) | 1.0178 (0.0963) | 0.6876 (0.0629) |

Table 5: Means of EM estimates and MADE (within parentheses), for TVZIP-INARCH (2) models where zero-inflation is driven by a sinusoidal function.

| Model | $N$ | $A$ | $B$ | $\alpha_0$ | $\alpha_1$ | $\alpha_2$ |
|---|---|---|---|---|---|---|
|  | **True values** | **0.10** | **0.10** | **1.00** | **0.20** | **0.20** |
| **B1** | 120 | 0.0875 (0.0556) | 0.0841 (0.0536) | 1.0529 (0.1601) | 0.1877 (0.0874) | 0.1765 (0.0863) |
| **B1** | 360 | 0.0955 (0.0317) | 0.0972 (0.0290) | 1.0239 (0.0945) | 0.1943 (0.0496) | 0.1906 (0.0503) |
|  | **True Values** | **-0.25** | **-0.25** | **2.00** | **0.30** | **0.20** |
| **B2** | 120 | -0.2485 (0.0430) | -0.2476 (0.0395) | 2.0524 (0.2563) | 0.2842 (0.0952) | 0.1906 (0.0931) |
| **B2** | 360 | -0.2514 (0.0234) | -0.2470 (0.0224) | 2.0254 (0.1478) | 0.2989 (0.0500) | 0.1901 (0.0536) |
|  | **True Values** | **-0.35** | **-0.30** | **1.00** | **0.40** | **0.30** |
| **B3** | 120 | -0.3480 (0.0453) | -0.2940 (0.0485) | 1.0358 (0.1712) | 0.3839 (0.1305) | 0.2781 (0.1324) |
| **B3** | 360 | -0.3510 (0.0282) | -0.2974 (0.0273) | 1.0149 (0.1004) | 0.3975 (0.0709) | 0.2914 (0.0762) |

Table 6: Means of EM estimates and MADE (within parentheses), for TVZIP-INGARCH (1, 1) models where zero-inflation is driven by a sinusoidal function.

| Model | N | A | B | $\alpha_0$ | $\alpha_1$ | $\beta_1$ |
|---|---|---|---|---|---|---|
| | **True values** | **0.10** | **0.10** | **1.00** | **0.20** | **0.20** |
| C1 | 120 | 0.0952 (0.0534) | 0.0898 (0.0558) | 0.9376 (0.2188) | 0.2218 (0.0793) | 0.2142 (0.1415) |
| | 360 | 0.1058 (0.0301) | 0.0983 (0.0296) | 0.9436 (0.1589) | 0.2292 (0.0556) | 0.2042 (0.1206) |
| | **True Values** | **-0.25** | **-0.25** | **2.00** | **0.30** | **0.20** |
| C2 | 120 | -0.2515 (0.0395) | -0.2466 (0.0404) | 1.7578 (0.3261) | 0.3616 (0.0972) | 0.1936 (0.1175) |
| | 360 | -0.2516 (0.0231) | -0.2500 (0.0221) | 1.8070 (0.2347) | 0.3843 (0.0884) | 0.1539 (0.0859) |
| | **True Values** | **-0.35** | **-0.30** | **1.00** | **0.40** | **0.30** |
| C3 | 120 | -0.3583 (0.0415) | -0.2931 (0.0453) | 0.9280 (0.2186) | 0.4671 (0.1288) | 0.2364 (0.1683) |
| | 360 | -0.3590 (0.0258) | -0.2984 (0.0264) | 0.9350 (0.1494) | 0.4961 (0.1086) | 0.2018 (0.1367) |

The above simulation results show that the MLE and EM procedures produced almost identical means of estimates for the parameters in TVZIP-INGARCH (1) and TVZIP-INARCH (2) models. For example, the means of estimates in Table 1 are almost identical to the corresponding means of estimates in Table 4. The MADE values for corresponding estimates are also almost identical across the two estimation methods. This similarity extends to means of corresponding parameter estimates across Tables 2 and 5 as well. Even in cases where the means of estimates are not identical, they are extremely close. For instance, for Model A3 with $N = 360$ and the true value of $\alpha_1$ equal to 0.70, the mean of the MLE estimates for this parameter is 0.6875 while the mean of the EM estimates is 0.6876.

However, for the TVZIP-INGARCH (1, 1) process, there are relatively larger differences between means of estimates for the MLE and EM methods. For example, when $N = 120$ with true parameter values of $\alpha_0 = 1.00$, $\alpha_1 = 0.20$, $\beta_1 = 0.20$, the means of parameters estimates are $\alpha_0 = 0.9420$, $\alpha_1 = 0.2226$, $\beta_1 = 0.2078$ for the MLE method (Table 3 row 1) while EM mean estimates are $\alpha_0 = 0.9376$, $\alpha_1 = 0.2218$, $\beta_1 = 0.2142$ (Table 6, row 1). Generally, the larger sample size produced means of estimates closer to their true value. An exception to this is observed in the case of TVZIP-INGARCH (1, 1), where the means of estimates for $(\alpha_1, \beta_1)$ did not improve with increasing sample size. For instance, when $N = 120$ with true values of $\alpha_1 = 0.40$, $\beta_1 = 0.30$, the means of the corresponding MLE estimates are 0.4666 and 0.2327 respectively (see Table 3, Model C3) , but when the sample size increase to 360, the means of the corresponding estimates changed to 0.4958 and 0.2022 respectively, which is a movement in the wrong direction. The MADE values decreased consistently with increasing sample size for both the MLE and EM estimation methods. Another relevant observation is that the simulation results for the INGARCH portion $(\alpha_1, \beta_1)$ of the model behave similar to the results obtained by Zhu (2012a) in the sense that the mean of the estimates are not very close to the true values even with higher sample sizes.

### 4.2. SIMULATION STUDY FOR CASE 2: ZERO-INFLATION FUNCTION DRIVEN BY AN EXOGENOUS VARIABLE

In this part of the study, we allow the exogenous variable to generate zeros through a logistic model as described in Equation (2.5). The parameter vector for the simulation study under this scenario is $\Phi = (\delta_0, \delta_1, \alpha_0, \alpha_1, \alpha_2, \beta_1)$, where $\delta_0$ and $\delta_1$ are the parameters in the logistic part of the model, while $(\alpha_0, \alpha_1), (\alpha_0, \alpha_1, \alpha_2)$, and $(\alpha_0, \alpha_1, \beta_1)$ are the parameter combination for TVZIP-INARCH (1), TVZIP-INGARCH (2), and TVZIP-INGARCH (1,

1) models, respectively. The parameter combination $\delta_0$ and $\delta_1$ were set to (-2, 0), (-1, -1) and (2, 1), representing three types of changes in zero-inflation probability with respect to the exogenous variable. These are no change ($\delta_1 = 0$), decrease ($\delta_1 < 0$), and increase ($\delta_1 > 0$) in the zero-inflation probability with increasing values of the exogenous variable. We generated an exogenous stationary AR (12) time series given in Equation (2.5) using $\eta = 0.25$. The following models were considered:

**(A)** TVZIP-INARCH (1) models: $\Phi = (\delta_0, \delta_1, \alpha_0, \alpha_1)$

    A1.    (-2.00, 0.00, 1.00, 0.40)

    A2.    (-1.00, -1.00, 2.00, 0.50)

    A3.    (2.00, 1.00, 1.00, 0.70)

**(B)** TVZIP-INARCH (2) models: $\Phi = (\delta_0, \delta_1, \alpha_0, \alpha_1, \alpha_2)$

    B1.    (-2.00, 0.00, 1.00, 0.20, 0.20)
    B2.    (-1.00, -1.00, 2.00, 0.30, 0.20)
    B3.    (2.00, 1.00, 1.00, 0.40, 0.30)

**(C)** TVZIP-INGARCH (1,1) models: $\Phi = (\delta_0, \delta_1, \alpha_0, \alpha_1, \beta_1)$

    C1.    (-2.00, 0.00, 1.00, 0.20, 0.20)
    C2.    (-1.00, -1.00, 2.00, 0.30, 0.20)
    C3.    (2.00, 1.00, 1.00, 0.40, 0.30)

Tables 7 through 9 provide the simulation results for the MLE estimation techniques, while Tables 10 through 12 provide EM algorithm related simulation results.

Table 7: Means of MLE Estimates and MADE (within parentheses), for TVZIP-INARCH (1) models where zero-inflation is driven by an exogenous variable.

| Model | $N$ | $\delta_0$ | $\delta_1$ | $\alpha_0$ | $\alpha_1$ |
|---|---|---|---|---|---|
| | **True values** | **-2.00** | **0.00** | **1.00** | **0.40** |
| **A1** | 120 | -2.4301 (0.7134) | -0.0331 (0.5515) | 1.0169 (0.1381) | 0.3808 (0.0837) |
| | 360 | -2.0982 (0.3258) | -0.0162 (0.2559) | 1.0143 (0.0817) | 0.3892 (0.0487) |
| | **True Values** | **-1.00** | **-1.00** | **2.00** | **0.50** |
| **A2** | 120 | -1.0467 (0.2382) | -1.0711 (0.2580) | 2.0193 (0.2143) | 0.4864 (0.0836) |
| | 360 | -1.0209 (0.1338) | -1.0182 (0.1400) | 2.0073 (0.1199) | 0.4932 (0.0476) |
| | **True Values** | **2.00** | **1.00** | **1.00** | **0.70** |
| **A3** | 120 | 2.0047 (0.4342) | 1.1592 (0.4153) | 0.9914 (0.2978) | 0.4605 (0.3699) |
| | 360 | 1.9976 (0.2197) | 1.0365 (0.1947) | 1.0037 (0.1694) | 0.5662 (0.2739) |

Table 8: Means of MLE Estimates and MADE (within parentheses), for TVZIP-INARCH (2) models where zero-inflation is driven by an exogenous variable.

| Model | $N$ | $\delta_0$ | $\delta_1$ | $\alpha_0$ | $\alpha_1$ | $\alpha_2$ |
|---|---|---|---|---|---|---|
| | **True values** | **-2.00** | **0.00** | **1.00** | **0.20** | **0.20** |
| **B1** | 120 | -2.4576 (0.7430) | -0.0369 (0.5504) | 1.0217 (0.1560) | 0.1947 (0.0741) | 0.1846 (0.0729) |
| | 360 | -2.1479 (0.3557) | 0.0024 (0.2779) | 1.0147 (0.0950) | 0.1956 (0.0474) | 0.1904 (0.0453) |
| | **True Values** | **-1.00** | **-1.00** | **2.00** | **0.30** | **0.20** |
| **B2** | 120 | -1.0505 (0.2390) | -1.0593 (0.2436) | 2.0354 (0.2494) | 0.2911 (0.0829) | 0.1930 (0.0721) |
| | 360 | -1.0132 (0.1258) | -1.0191 (0.1345) | 2.0092 (0.1418) | 0.2989 (0.0476) | 0.1977 (0.0453) |
| | **True Values** | **2.00** | **1.00** | **1.00** | **0.40** | **0.30** |
| **B3** | 120 | 1.9814 (0.4398) | 1.1959 (0.4396) | 0.9418 (0.2881) | 0.3127 (0.2752) | 0.2656 (0.2253) |
| | 360 | 1.9801 (0.2139) | 1.0532 (0.1973) | 0.9797 (0.1665) | 0.3584 (0.2285) | 0.2952 (0.1940) |

Table 9: Means of MLE Estimates and MADE (within parentheses), for TVZIP-INGARCH (1, 1) models where zero-inflation is driven by an exogenous variable.

| Model | $N$ | $\delta_0$ | $\delta_1$ | $\alpha_0$ | $\alpha_1$ | $\beta_1$ |
|---|---|---|---|---|---|---|
|  | **True values** | **-2.00** | **0.00** | **1.00** | **0.20** | **0.20** |
| **C1** | 120 | -2.1527 (0.4716) | -0.0183 (0.3284) | 0.9318 (0.2158) | 0.2257 (0.0790) | 0.2090 (0.1381) |
| | 360 | -2.0880 (0.3031) | 0.0025 (0.2342) | 0.9411 (0.1509) | 0.2280 (0.0512) | 0.1887 (0.1070) |
|  | **True Values** | **-1.00** | **-1.00** | **2.00** | **0.30** | **0.20** |
| **C2** | 120 | -1.0226 (0.2184) | -1.0412 (0.2332) | 1.7477 (0.3285) | 0.3924 (0.1142) | 0.1620 (0.1008) |
| | 360 | -1.0083 (0.1244) | -1.0131 (0.1340) | 1.8084 (0.2163) | 0.3992 (0.1021) | 0.1337 (0.0869) |
|  | **True Values** | **2.00** | **1.00** | **1.00** | **0.40** | **0.30** |
| **C3** | 120 | 1.9804 (0.3863) | 1.0538 (0.2881) | 0.8406 (0.3229) | 0.3927 (0.3003) | 0.2271 (0.2395) |
| | 360 | 1.9927 (0.2193) | 1.0375 (0.1920) | 0.8296 (0.2337) | 0.4782 (0.2648) | 0.2271 (0.2242) |

Table 10: Means of EM Estimates and MADE (within parentheses), for TVZIP-INARCH (1) models where zero-inflation is driven by an exogenous variable.

| Model | $N$ | $\delta_0$ | $\delta_1$ | $\alpha_0$ | $\alpha_1$ |
|---|---|---|---|---|---|
| | **True values** | **-2.00** | **0.00** | **1.00** | **0.40** |
| **A1** | 120 | -2.5278 (0.8110) | -0.0402 (0.5931) | 1.0160 (0.1386) | 0.3806 (0.0837) |
| | 360 | -2.1031 (0.3307) | -0.0154 (0.2575) | 1.0144 (0.0817) | 0.3892 (0.0487) |
| | **True Values** | **-1.00** | **-1.00** | **2.00** | **0.50** |
| **A2** | 120 | -1.0467 (0.2382) | -1.0711 (0.2580) | 2.0193 (0.2143) | 0.4864 (0.0836) |
| | 360 | -1.0209 (0.1338) | -1.0182 (0.1400) | 2.0073 (0.1198) | 0.4932 (0.0476) |
| | **True Values** | **2.00** | **1.00** | **1.00** | **0.70** |
| **A3** | 120 | 2.0083 (0.4307) | 1.1504 (0.4065) | 0.9923 (0.2968) | 0.4606 (0.3699) |
| | 360 | 1.9976 (0.2196) | 1.0365 (0.1947) | 1.0037 (0.1694) | 0.5662 (0.2739) |

Table 11: Means of EM Estimates and MADE (within parentheses), for TVZIP-INARCH (2) models where zero-inflation is driven by an exogenous variable.

| Model | $N$ | $\delta_0$ | $\delta_1$ | $\alpha_0$ | $\alpha_1$ | $\alpha_2$ |
|---|---|---|---|---|---|---|
| | **True values** | **-2.00** | **0.00** | **1.00** | **0.20** | **0.20** |
| **B1** | 120 | -2.1663 (0.4741) | -0.0180 (0.3259) | 1.0296 (0.1554) | 0.1957 (0.0746) | 0.1852 (0.0730) |
| | 360 | -2.0915 (0.3012) | 0.0072 (0.2335) | 1.0165 (0.0946) | 0.1958 (0.0474) | 0.1907 (0.0453) |
| | **True Values** | **-1.00** | **-1.00** | **2.00** | **0.30** | **0.20** |
| **B2** | 120 | -1.0505 (0.2390) | -1.0593 (0.2436) | 2.0354 (0.2494) | 0.2911 (0.0829) | 0.1930 (0.0721) |
| | 360 | -1.0132 (0.1258) | -1.0191 (0.1345) | 2.0092 (0.1418) | 0.2989 (0.0476) | 0.1977 (0.0453) |
| | **True Values** | **2.00** | **1.00** | **1.00** | **0.40** | **0.30** |
| **B3** | 120 | 1.9865 (0.4348) | 1.1890 (0.4326) | 0.9420 (0.2878) | 0.3137 (0.2752) | 0.2649 (0.2248) |
| | 360 | 1.9801 (0.2139) | 1.0532 (0.1973) | 0.9797 (0.1655) | 0.3585 (0.2285) | 0.2944 (0.1936) |

Table 12: Means of EM Estimates and MADE (within parentheses), for TVZIP-INGARCH (1, 1) models where zero-inflation is driven by an exogenous variable.

| Model | $N$ | $\delta_0$ | $\delta_1$ | $\alpha_0$ | $\alpha_1$ | $\hat{\beta}_1$ |
|---|---|---|---|---|---|---|
| | **True values** | **-2.00** | **0.00** | **1.00** | **0.20** | **0.30** |
| C1 | 120 | -2.1170 (0.4717) | -0.0190 (0.3224) | 0.9289 (0.2108) | 0.2251 (0.0794) | 0.2120 (0.1371) |
| C1 | 360 | -2.1166 (0.3504) | 0.0029 (0.2631) | 0.9347 (0.1536) | 0.2275 (0.0514) | 0.1925 (0.1082) |
| | **True Values** | **-1.00** | **-1.00** | **2.00** | **0.30** | **0.20** |
| C2 | 120 | -1.0236 (0.2191) | -1.0413 (0.2330) | 1.7423 (0.3318) | 0.3924 (0.1141) | 0.1612 (0.0994) |
| C2 | 360 | -1.0083 (0.1244) | -1.0131 (0.1340) | 1.8084 (0.2163) | 0.3992 (0.1021) | 0.1337 (0.0869) |
| | **True Values** | **2.00** | **1.00** | **1.00** | **0.40** | **0.30** |
| C3 | 120 | 1.9798 (0.3665) | 1.0452 (0.2845) | 0.8236 (0.3205) | 0.3963 (0.3005) | 0.2297 (0.2359) |
| C3 | 360 | 1.9871 (0.2187) | 1.0339 (0.1909) | 0.8092 (0.2478) | 0.4805 (0.2631) | 0.2202 (0.2147) |

Both the MLE and EM methods produced fairly accurate estimates for the parameters across both TVZIP-INARCH($p$) models. However, for the TVZIP-INGARCH (1, 1) case where the GARCH portion of the parameters $(\alpha_1, \beta_1)$ show relatively less accurate estimates, for both procedures even for the large sample size case. Increasing sample size does not seem to rectify this relative inaccuracy. For instance, when $N = 120$ with true values of $\alpha_1 = 0.40$ and $\beta_1 = 0.30$ the means of the corresponding parameter estkimates are $0.3927$ and $0.2271$, respectively, for the MLE method, and when $N = 360$ these means of estimates change to 0.4782 and 0.2271, respectively. A similar phenomenon is seen in the simulation results of Zhu (2012a). In general, MADE values decrease with increase in sample size.

## 5. REAL DATA EXAMPLE

In this section the proposed TVZIP-INGARCH (*p*, *q*) model is applied to two real-world dataset and the performance it is compared to the model proposed by Zhu (2012a). The Akaike Information Criterion (AIC) and the Bayesian Information Criterion (BIC) were employed to select the best model among the collection of competing models. The first example is based on the *Influenza A associated pediatric deaths* data set downloaded from the Center for Disease Control and Prevention webpage. This data was modeled by using the TVZIP-INGARCH (*p*, *q*) formulation a with sinusoidal zero-inflation function. In the second example, we used the *Pediatric mortality caused by Influenza B* data set downloaded from the same webpage to demonstrate the performance of TVZIP-INGARCH (*p*, *q*) process where the zero-inflation is driven by an exogenous variable.

### 5.1. REAL DATA EXAMPLE - USE OF A DETERMINISTIC SINUSOIDAL ZERO-INFLATION FUNCTION

The TVZIP-INARCH (1), TVZIP-INARCH (2), and TVZIP-INGARCH (1, 1) models with sinusoidal zero-inflation was fitted to the *Influenza A associated pediatric mortality* data set. Results were compared to those from fitting the ZIP_INGARCH model of Zhu (2012a). The data set provides weekly count data of U.S. pediatric deaths caused by Type A influenza viruses over the time period from week 40 of 2015 to week 43 of year 2018. This consists of 160 weekly observations of pediatric death counts. The data were extracted from weekly U.S. Influenza Surveillance report, which was published by Center for Disease Control and Prevention (CDC) ([Influenza-associated Pediatric Mortality (cdc.gov)](Influenza-associated Pediatric Mortality (cdc.gov))). A copy of the data set can be made available by the corresponding author upon request. Summary statistics of the data showed a mean of 1.506 and a variance of 6.352, suggestive of over dispersion. Figure 1 illustrates the frequency of each pediatric mortality case caused by virus type A using a bar chart. Observe that there are 86 zeros, which comprises 53.8% of the total time points.

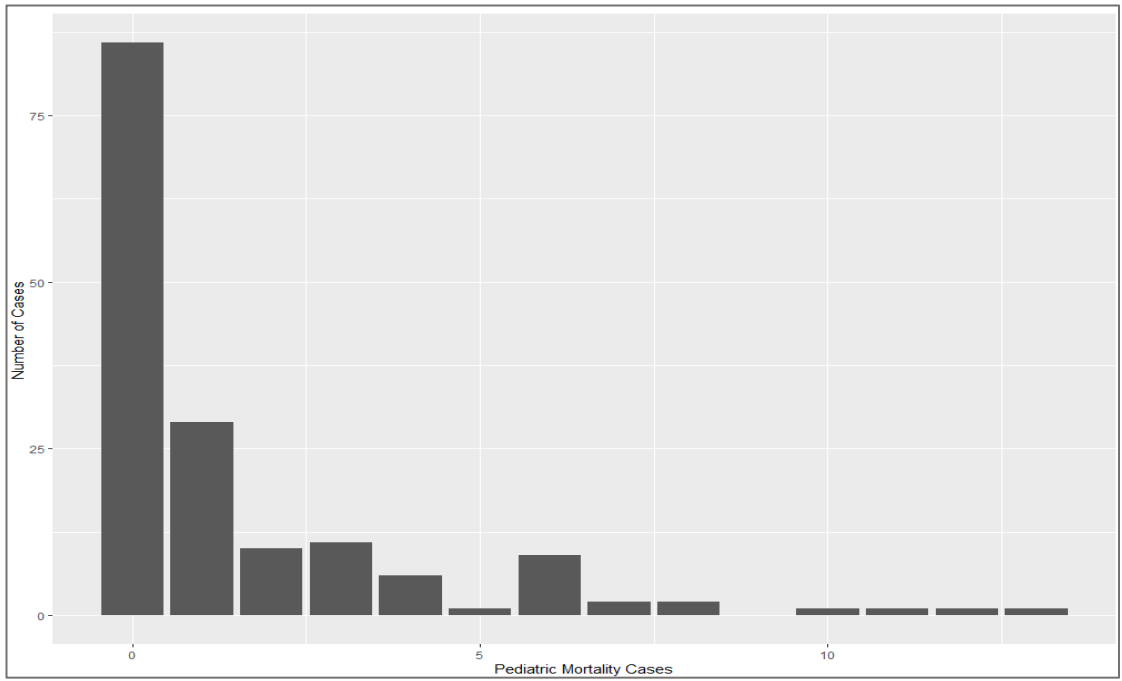

Figure 1: Bar chart of the pediatric death counts caused by virus Type A.

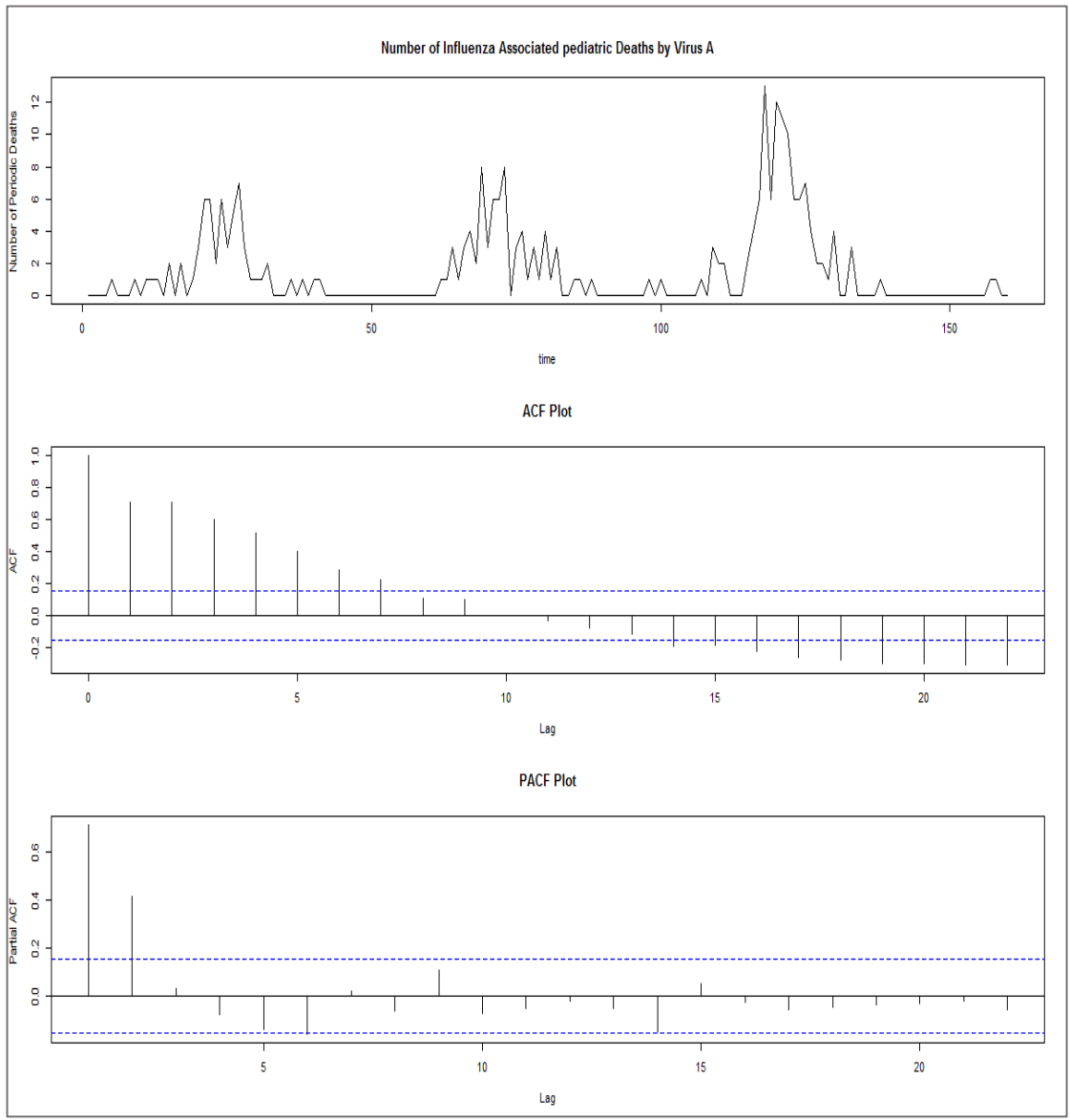

Figure 2: Pediatric mortality time series plot, sample auto covariance plot and sample partial auto covariance plot.

Figure 2 illustrates the original time series of the counts followed by the plots of its autocorrelation function (ACF) and the partial autocorrelation function PACF, respectively. The bar chart shows the excess number of zeros in the data, and the time series plot demonstrates prolonged periods of zero counts, which supports the use of zero-inflated poisson time series models to analyze this data. Furthermore, we can observe an annual seasonality in the portions of the time series with excessive zero counts.

To understand the zero-inflation behaviour of this data set, we aggregated weekly data in to its corresponding calendar month and constructed the total monthly zero mortality counts. These counts were then converted into a monthly proportion by dividing each monthly total zeros by the maximum zero count over the observed period to scale values at or below one. The plot of the monthly proportion of zero counts exhibits an approximate sinosidual behaviour throughout the observed time span. Thus, the general sinosidual function mentioned in Equation 2.4 can be used to model the zero-inflation behaviour of this data. More details of this modeling will be describe later in this section.

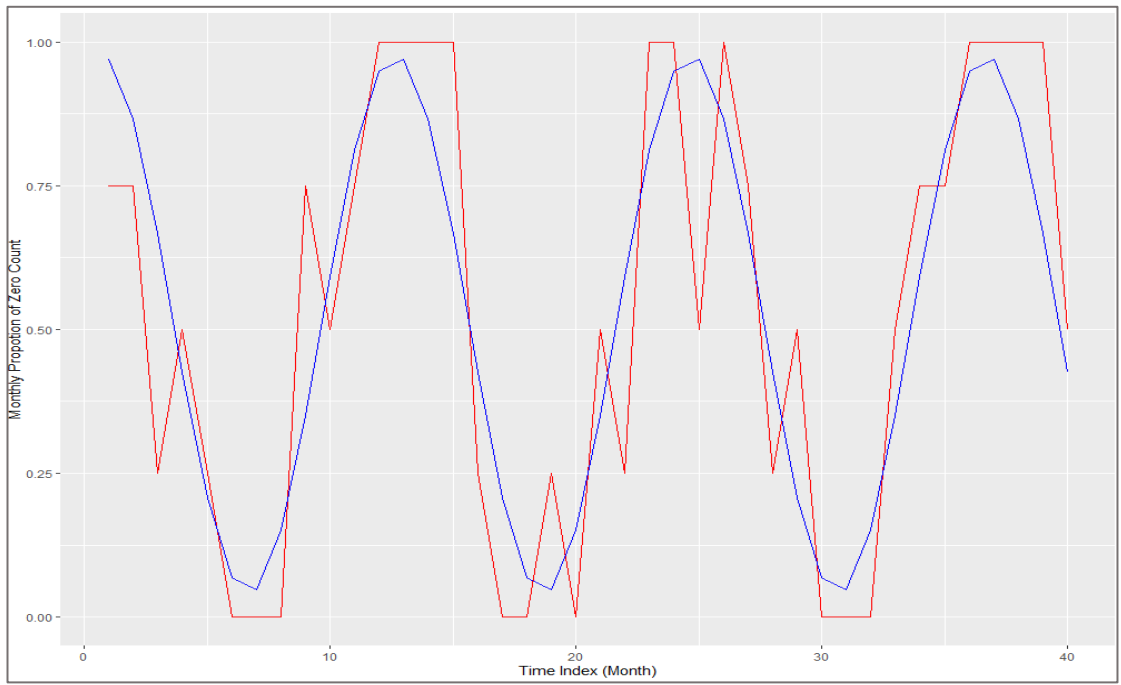

Figure 3: Monthly proportion of zero mortality counts (red) versues the fitted sinusoidal zero-inflated function (blue).

Three time series models were fitted namely; TVZIP-INARCH(1) (*M1*), TVZIP-INARCH (2) (*M2*) and TVZIP-INGARCH (1,1) (*M3*). In addition, there deterministic time varying zero-inflation fictions were empoyed for each of the above models. These senarios are labled *S1*, *S2*, and *S3*. In senario *S1*, it was assumed that zero-inflation probability is a constant over time as assumed by Zhu (2012a). In the model listed as *S2*, we assumed a piecewise constant zero-inflation fucntion. The monthly zero-inflation was allowed to vary according to a sinusoidal zero-inflation function with a 12-month priod, but then the weeks within a given month were assigned the same zero-inflation value associated with that month. The models under *S3* used a sinusoidal function describing a time varying zero-inflation probability that changes from week to week. The EM algorithem were used to estimate the model parameters and the results were recorded in the Table 13. The EM algorithm was chosen to keep in line with the method used in Zhu (2012a) for its real data example. We used both AIC and BIC model selection criteria to identify the best fitting model.

Table 13: Estimated parameters, AIC and BIC for the pediatric death counts cause by virus A.

| Model | $\omega$ | $A$ | $B$ | $\alpha_0$ | $\alpha_1$ | $\alpha_2$ | $\beta_1$ | AIC | BIC |
|---|---|---|---|---|---|---|---|---|---|
| S1M1 | 0.1806 | | | 0.4714 | 0.8416 | | | 451.6289 | 460.8544 |
| S1M2 | 0.1012 | | | 0.2539 | 0.4967 | 0.4187 | | 419.8733 | 432.1740 |
| S1M3 | 0.0909 | | | 0.0807 | 0.4804 | | 0.5057 | 420.6112 | 432.9119 |
| S2M1 | | 0.1023 | 0.4467 | 1.1030 | 0.7165 | | | 411.1064 | 423.4071 |
| **S2M2** | | 0.0775 | 0.4266 | 0.7446 | 0.4715 | 0.3595 | | **397.4778** | **412.8537** |
| S2M3 | | 0.0733 | 0.4219 | 0.3730 | 0.4832 | | 0.4413 | 401.4099 | 416.7857 |
| S3M1 | | -0.3116 | 0.3214 | 1.0292 | 0.7193 | | | 420.2297 | 432.5304 |
| S3M2 | | -0.3074 | 0.2613 | 0.6460 | 0.4544 | 0.3927 | | 402.9449 | 418.3208 |
| S3M3 | | -0.3009 | 0.2420 | 0.2726 | 0.4698 | | 0.4773 | 407.6126 | 422.9885 |

Based on results in Table 13, models with cyclically varying zero-inflation function had lower AIC and BIC values in general, when compared to the results for models with constant zero-inflation. In other words, models under the *S2* and *S3* fit the data better than the models under *S1*. Among all the TVZIP-INARCH (1), TVZIP-INARCH (2) and TVZIP-INGARCH (1, 1) models, TVZIP-INARCH (2) process had lower AIC and BIC values in all the three scenarios. Finally, based on both information criteria, the TVZIP-INARCH (2) model with zero-inflation model *S2,* provided the best fit to the data. In this model, we assumed that weeks within any given month has a constant zero-inflation, yet monthly zero-inflation varies cyclically.

## 5.2. REAL DATA EXAMPLE - ZERO-INFLATION FUNCTION IS DRIVEN BY EXOGENOUS VARIABLE

In this sub section we examine the performance of the TVZIP-INGARCH ($p$, $q$) models where the zero-inflated function was driven by an exogenous variable. *Influenza B associated pediatric mortality* data set was used, and the TVZIP-INARCH (1), TVZIP-INARCH (2) and TVZIP-INGARCH (1, 1) models were fitted to the data. We selected the weekly average of nationwide low temperatures as the exogenous variables that drives the zero-inflation probability. This is motivated by one of the reasons generally accepted as a cause for the easy transmission of influenza during winter months is the cold temperatures driving people indoors. Influenza B associated pediatric mortality data set was accessed from the weekly U.S. Influenza Surveillance Report ([Influenza-associated Pediatric Mortality (cdc.gov)](#)), and the temperature data were extracted from the United States National Oceanic and Atmospheric Administration (NOAA) weather prediction center (https://www.wpc.ncep.noaa.gov). Both data sets spanned the time period from week 40 in year 2014 to week 39 in year 2018. The influenzas B data set contained 209 weekly observations of pediatric death counts. Copies of both data sets can be made available by the corresponding author upon request. Summary statistics of infant mortality cases due to influenza B showed a mean of 0.8517 and a variance of 2.4538. Since the empirical variance was higher than the empirical mean, the data exhibits over dispersion. Figure 4 illustrates the frequency of each pediatric mortality count caused by virus Type B using a bar chart. The bar chart shows that there are 128 zeros, which comprises 61.2% of total of the time points. This suggest that the pediatric mortality data are zero-inflated.

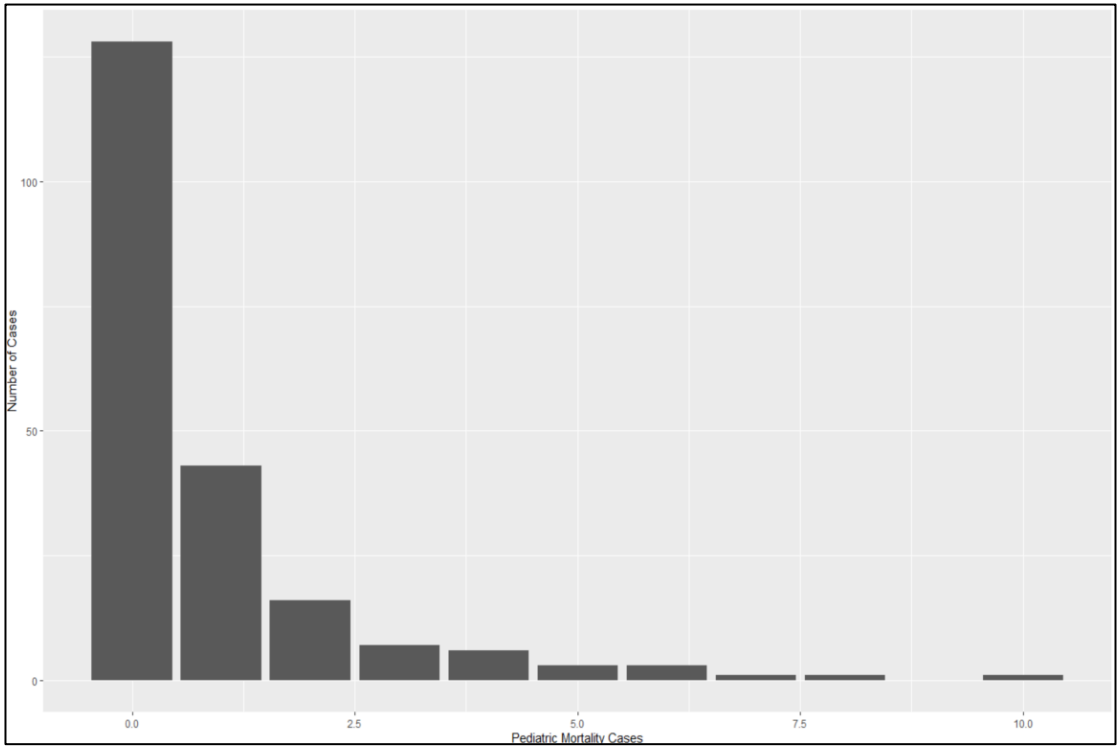

Figure 4: Bar hart of the pediatric death counts caused by virus type "A".

The time series plot, ACF, and PACF plots are given in Figure 5. Based on time series plot, we can see that there is an annual seasonality exhibited in this data set. Moreover, it shows that there were periods with clusters of zeros that occur periodically. Therefore, as discussed in Section 5.1, we suggested TVZIP-INGARCH ($p$, $q$) to model the count data series.

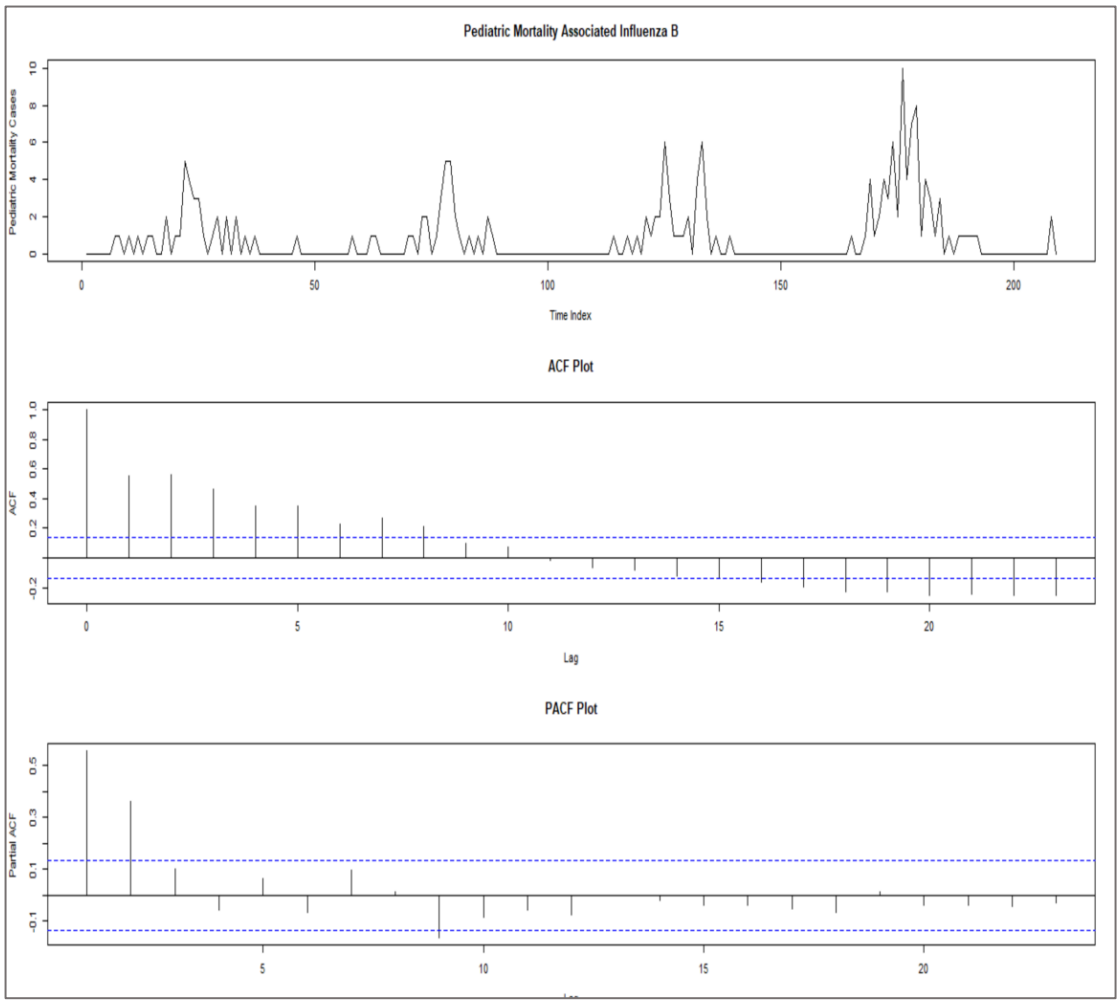

Figure 5: Pediatric mortality time series plot, sample auto covariance plot and sample partial auto covariance plot.

In this example, the time varying zero-inflation was modeled by considering an exogenous time series, namely the weekly average of nationwide low temperature. Comparison of the two time series plots is given in Figure 6.

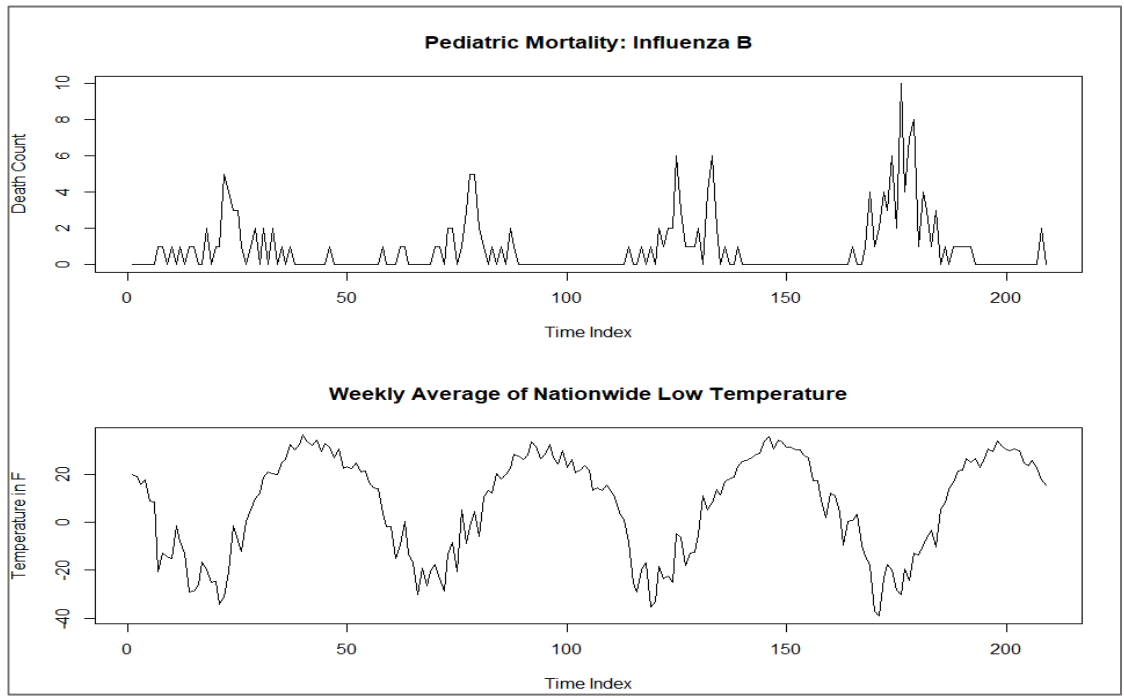

Figure 6: Time series plots of pediatric mortality counts (upper panel) and weekly average of nationwide low temperature.

Figure 6 shows that periods with higher values of low temperature coincide with periods of zero pediatric mortality caused by Influenza B. Hence, it was suggestive that periods with high zero counts (low pediatric mortality) are related to periods with higher values of low temperature. Thus, we used the weekly average low temperature data to model the time varying zero-inflated component in the pediatric mortality data set. In this example, time varying zero-inflation process was modeled using the formulation from Equation 2.5.

We fitted three different time series models: TVZIP-INARCH(1) (*M1*), TVZIP-INARCH (2) (*M2*), and TVZIP-INGARCH (1,1) (*M3*) for two different senarios ( *S1* and *S2* ) . In senario *S1,* we considered that there is a constant zero-inflation probabilty throughout the observation period, as assumed by Zhu (2012a). For the models listed under *S2,* we assumed there is a time varying zero-inflation function and we modeled it by using a logistic regression with low temperature as the independent variable.The EM algorithm was used to estimate the model parameters, as was done by Zhu (2012a) and the results are recorded in Table 14. We used both AIC and BIC model selection criterion to identify the best fit for the data.

Table 14: Estimated parameters, AIC and BIC for the pediatric death counts cause by virus B.

| Model | $\omega$ | $\delta_0$ | $\delta_1$ | $\alpha_0$ | $\alpha_1$ | $\alpha_2$ | $\beta_1$ | AIC | BIC |
|---|---|---|---|---|---|---|---|---|---|
| S1M1 | 0.2183 | | | 0.4517 | 0.6746 | | | 488.2955 | 498.3225 |
| S1M2 | 0.0575 | | | 0.1737 | 0.4095 | 0.4454 | | 442.2950 | 455.6644 |
| S1M3 | 0.1478 | | | 0.0001 | 0.3766 | | 0.7631 | 441.4626 | 454.8319 |
| S2M1 | | -1.7436 | 0.1115 | 0.7258 | 0.5789 | | | 453.1119 | 466.4812 |
| **S2M2** | | -2.5485 | 0.1190 | 0.3478 | 0.4027 | 0.4101 | | **427.7886** | **444.5003** |
| S2M3 | | -1.6346 | 0.1048 | 0.2385 | 0.3591 | | 0.8305 | 428.6728 | 445.3844 |

Table 14 shows the models that fall under the *S2* exhibited low AIC and BIC values compared to the models under *S1*. Thus, the use of a time-varying zero-inflation model improved the fit and our use of an exogenous variable to model the time varying zero-inflation function produced reasonable results. It is possible that functions based on other exogenous variables may have done a better job, but the main objective of this exercise is to illustrate the utility of the methodology we have proposed in modeling a real-life situation. Based on the AIC and BIC values, TVZIP-INARCH (2) with time varying zero-inflation function provided a better fit to the pediatric mortality data when compared to other models considered in this study.

# 6. CONCLUSIONS

A time varying zero-inflated Poisson integer GARCH (TVZIP-INGARCH) model was proposed to accommodate situations where zero-inflation is driven by either a deterministic function of time or a set of exogenous variables. Monte-Carlo simulation results indicate that the Expected Maximization (EM) and Maximum Likelihood Estimation (MLE) methods produce very similar results with respect to parameter estimates. It is observed that both EM and MLE techniques estimated the model parameters with good accuracy when the underlying model has a purely ARCH component. When the model has a GARCH component, the GARCH parameters are estimated with lesser accuracy, which is a phenomenon also seen in the a study of the existing ZIP-INGARCH model proposed by Zhu (2012a). When tested on two real-life data sets, the TVZIP-INGARCH models performed better than the ZIP-INGARCH formulation, illustrating the utility of the proposed model. In addition, the flexibility of modeling zero-inflation through deterministic cyclical functions or through exogenous time series provide the proposed model added versatility.

**APPENDIX**

In this appendix we derive some of the results mentioned in the text of the article. Rest of the appendix is organized as follows. The conditional mean and the conditional variance of $X_t | \mathbb{F}_{t-1}$ are derived in Appendix A.1. Expression for the conditional mass function for the $X_t | Z_t, \mathbb{F}_{t-1}$ is given in Appendix A.2. Finally, Appendix B specifies and proves the conditions that need to be satisfied to ensure $\omega_t \in (0,1)$ in the sinusoidally varying zero-inflation case.

**Appendix A.**

**A.1 Derivation of the Conditional Mean**

Let $\{X_t | t \in \mathbb{Z}\}$ be a discrete time series of count data and conditional distribution of $X_t | \mathbb{F}_{t-1} \sim ZIP(\lambda_t, \omega_t)$ as described in (2.1). The ZIP process has two types of zeros namely, zeroes coming from the Poisson distribution with the rate parameter $\lambda_t$ and the zeros that are generated by an independent process. Let us define $\{Z_t | t \in \mathbb{Z}\}$ to be a time series of Bernoulli random variable such that $Z_t = 0$ if $X_t$ arises from the Poisson distribution and $Z_t = 1$ if $X_t = 0$ and is generated by the independent process. Therefore $Z_t | \mathbb{F}_{t-1} \sim \text{Bernoulli}(\omega_t)$ such that,

$$P(Z_t = z | \mathbb{F}_{t-1}) \begin{cases} \omega_t : z = 1. \\ 1 - \omega_t : z = 0. \end{cases}$$

Let define the conditional probability mass function of $X_t | Z_t, \mathbb{F}_{t-1} \sim \text{Poisson}((1-Z_t)\lambda_t)$. Hence the conditional distribution of $X_t | \mathbb{F}_{t-1} \sim ZIP(\lambda_t, \omega_t)$ can be expressed as:

$$P(X_t = k | \mathbb{F}_{t-1}) = \sum_{z=0}^{1} P(X_t = k | Z_t, \mathbb{F}_{t-1}) P(Z_t = z | \mathbb{F}_{t-1}),$$

The Conditional expectation of $X_t | \mathbb{F}_{t-1}$ is:

$$E(X_t | \mathbb{F}_{t-1}) = E\big[E(X_t | Z_t, \mathbb{F}_{t-1})\big],$$

$$= E\big[(1-Z_t)\lambda_t | \mathbb{F}_{t-1}\big],$$
$$= \lambda_t E\big[(1-Z_t) | \mathbb{F}_{t-1}\big],$$
$$= \lambda_t (1-\omega_t).$$

The Conditional variance of $X_t | \mathbb{F}_{t-1}$ is:

$$Var(X_t | \mathbb{F}_{t-1}) = Var\big[E(X_t | Z_t, \mathbb{F}_{t-1})\big] + E\big[Var(X_t | Z_t, \mathbb{F}_{t-1})\big],$$
$$Var\big[E(X_t | Z_t, \mathbb{F}_{t-1})\big] = Var\big[(1-Z_t)\lambda_t | \mathbb{F}_{t-1}\big] = \lambda_t^2 (1-\omega_t)\omega_t,$$
$$E\big[Var(X_t | Z_t, \mathbb{F}_{t-1})\big] = E\big[(1-Z_t)\lambda_t | \mathbb{F}_{t-1}\big] = \lambda_t (1-\omega_t),$$
$$Var(X_t | \mathbb{F}_{t-1}) = \lambda_t^2 (1-\omega_t)\omega_t + \lambda_t (1-\omega_t) = \lambda_t (1-\omega_t)(1+\lambda_t \omega_t).$$

**A.2 Derivation of the Conditional Mass function**

The conditional probability mass function of $Z_t | \mathbb{F}_{t-1}$ is given by,

$$P(Z_t = z_t) = \omega_t^{z_t} (1-\omega_t)^{1-z_t}.$$

The conditional probability mass function of $X_t | Z_t, \mathbb{F}_{t-1}$ [ ~ Poisson$((1-Z_t)\lambda_t)$] is given by,

$$P(X_t = x_t | Z_t = z_t, \mathbb{F}_{t-1}) = z_t + (1-z_t)\frac{\lambda_t e^{-\lambda_t}}{x_t!} = \left(\frac{\lambda_t e^{-\lambda_t}}{x_t!}\right)^{(1-z_t)}.$$

Therefore, the conditional log likelihood function of $P(Z_t | \mathbb{F}_{t-1}).P(X_t | Z_t, \mathbb{F}_{t-1})$ is $L(\Phi) =$

$$\prod_{\forall t} P(Z_t | \mathbb{F}_{t-1}).P(X_t | Z_t, \mathbb{F}_{t-1}) = \prod_{\forall t} \omega_t^{z_t} (1-\omega_t)^{(1-z_t)} \left(\frac{\lambda_t e^{-\lambda_t}}{x_t!}\right)^{(1-z_t)} = \prod_{\forall t} \omega_t^{z_t} \left((1-\omega_t)\frac{\lambda_t e^{-\lambda_t}}{x_t!}\right)^{(1-z_t)}.$$

## Appendix B. Derivation of conditions for Zero-inflation Probability to lie inside (0, 1)

Let us define sinusoidal zero-inflation function $\omega_t = g(\mathbf{V}_t, \underline{\Gamma}) \in (0,1)$ as given bellow.

$$\omega_t = g(\mathbf{V}_t, \underline{\Gamma}) = A\sin\left(\frac{2\pi}{s}t\right) + B\cos\left(\frac{2\pi}{s}t\right) + C,$$

where $C = \sqrt{A^2 + B^2} + \delta$ and $\sqrt{A^2 + B^2} \leq \frac{1}{2} - \delta$ where $|A| \leq \frac{1}{2} - \delta, |B| \leq \frac{1}{2} - \delta$, and $\delta$ is a fixed positive value such that $\delta \in \left(0, \frac{1}{2}\right)$. Here

We consider two possible cases, one in which $\omega_t = g(\mathbf{V}_t, \underline{\Gamma})$ is a constant over time and the alternative case where it is not.

**Case 1**: If $A^2 + B^2 = 0$,

Since $A, B \in \mathbb{R}$ and $A^2 + B^2 = 0$ implies $A = B = 0$.

Therefore, $C = \delta$, where $\delta \in \left(0, \frac{1}{2}\right)$ then,

$\omega_t = g(\mathbf{V}_t, \underline{\Gamma}) = C$, is a constant over the time span under consideration.

**Case 2:** If $A^2 + B^2 \neq 0$ then,

$$\omega_t = A\sin\left(\frac{2\pi}{S}t\right) + B\cos\left(\frac{2\pi}{S}t\right) + C,$$

$$= \sqrt{A^2 + B^2}\left(\frac{A}{\sqrt{A^2+B^2}}\sin\left(\frac{2\pi}{S}t\right) + \frac{B}{\sqrt{A^2+B^2}}\cos\left(\frac{2\pi}{S}t\right) + \frac{C}{\sqrt{A^2+B^2}}\right),$$

$$= \sqrt{A^2 + B^2}\left(\cos(\varphi)\sin\left(\frac{2\pi}{S}t\right) + \sin(\varphi)\cos\left(\frac{2\pi}{S}t\right) + \frac{C}{\sqrt{A^2+B^2}}\right),$$

$$= \sqrt{A^2 + B^2}\left(\sin\left(\frac{2\pi}{S}t + \varphi\right) + \frac{C}{\sqrt{A^2+B^2}}\right),$$

$$= \sqrt{A^2 + B^2}\left(h(t, S, A, B,) + \frac{C}{\sqrt{A^2+B^2}}\right).$$

Here, $\cos(\varphi) = \frac{A}{\sqrt{A^2+B^2}}$ and $h(t, S, A, B,) = \sin\left(\frac{2\pi}{S}t + \varphi\right) \in [-1, 1]$. Thus, for a for a fixed $A, B$ and $C$, we have

$$\max(\omega_t) = \sqrt{A^2 + B^2} + C = 2\sqrt{A^2 + B^2} - \delta \leq 1 - \delta < 1,$$

$$\min(\omega_t) = -\sqrt{A^2 + B^2} + C = \delta, \text{ where } \delta \in \left(0, \frac{1}{2}\right).$$

This implies that the zero-inflation function lines within the interval (0, 1).